\documentclass[aps,prb,preprint,showpacs]{revtex4}%
\usepackage{graphicx}
\usepackage{amssymb}
\usepackage{acronym}
\usepackage{amsmath}
\usepackage{amsfonts}%
\newcommand{\be}{\begin{equation}}
\newcommand{\ee}{\end{equation}}
\newcommand{\bea}{\begin{eqnarray}}
\newcommand{\eea}{\end{eqnarray}}

\begin{document}
\title{First principles calculation of spin-interactions and magnetic ground states of
Cr trimers on Au(111)}
\author{A. Antal$^{1}$, B. Lazarovits$^{3}$, L. Udvardi$^{1,2}$,
 L. Szunyogh$^{1}$, B. \'Ujfalussy$^{3}$
and P. Weinberger$^{4}$ 
}
\affiliation{$^{1}$Department of Theoretical Physics ,
 Budapest University of Technology and Economics,
Budafoki \'{u}t 8, H-1111, Budapest, Hungary \linebreak
$^{2}$BME-HAS Group of Solid State Physics, Budapest University of Technology and Economics,
Budafoki \'{u}t 8, H-1111, Budapest, Hungary \linebreak
$^{3}$Research Institute for Solid State Physics and Optics,
Hungarian Academy of Sciences \\ H-1525 Budapest, PO Box 49, Hungary 
\linebreak
$^{4}$Center for
Computational Materials Science, Technical University Vienna, Gumpendorfer
Str. 1a, A-1060 Vienna, Austria
}
\date{\today}

\begin{abstract}
We present calculations of the magnetic ground states of Cr trimers in
different geometries on top of a Au(111) surface.
By using a least square fit method based on 
a fully relativistic embedded-cluster Green's function method  
first we determined the parameters of a classical vector-spin model 
consisting of second and fourth order interactions.
The newly developed method requires no symmetry constraints, therefore, 
it is throughout applicable for small nanoparticles of arbitrary geometry. 
The magnetic ground states were then found 
by solving the Landau-Lifshitz-Gilbert equations.
In all considered cases the configurational
energy of the Cr trimers is dominated 
by large antiferromagnetic nearest neighbor interactions, whilst
biquadratic spin-interactions have the second largest contributions
to the energy.
We find that an equilateral Cr trimer exhibits a frustrated
120$^\circ$ N\'eel type of ground state with a small out-of-plane component 
of the magnetization and we show that the Dzyaloshinsky-Moriya 
interactions determine the chirality of the magnetic ground state.
In cases of a linear chain and an isosceles trimer 
collinear antiferromagnetic ground states are obtained
with a magnetization lying parallel to the surface.

\end{abstract}

\maketitle

%%%%%%%%%%%%%%%%
%INTRODUCTION %
%%%%%%%%%%%%%%%%

\section{Introduction}
\label{sec:intro}

The development of nanoscale devices based on electron spin requires both a fundamental 
understanding of magnetic interactions and practical solutions to a variety of challenges.
Deposited clusters are of special interest due to their possible application in 
miniaturized data storage technology.
%Low dimensionality and interactions with the substrate result often in dramatic changes
%f the magnetic moment and the magnetic anisotropy energy as compared to the 
%orresponding bulk systems. 
The development of scanning tunneling microscopy (STM) and the ability to build clusters 
with well-controlled structures permit the 
measurement of various effects induced by local interactions within magnetic nanoclusters.  
Recent STM studies have investigated the coupling between the magnetic and 
electronic degrees of freedom of nanoparticles and the conducting substrate for 
adatoms \cite{monomer1,monomer2,monomer3}, dimers \cite{dimer1,dimer2} and trimers \cite{trimer}.
Very recently Wahl and coworkers \cite{wahl} have been able to estimate
the exchange coupling between Co atoms on Cu(001) surface by probing the Kondo resonance 
in terms of low temperature scanning tunneling spectroscopy.
A large number of theoretical efforts has been focused on the description of the Kondo 
effect of single atoms or small clusters.
\cite{kondo1,kondo2,kondo3,kondo4,kondo5} 
%strongly influenced by the interaction between the adatoms and with the substrate.

First principles studies of supported clusters are often useful for a clear interpretation 
of experimental results and can help a lot in understanding the underlying physical phenomena. 
Determining the, in general, non-collinear magnetic ground states of finite nanoparticles on an
ab initio level is clearly a demanding task of computational science. One stream of such works
is based on a fully unconstrained local spin-density approach (LSDA) implemented 
via the full-potential linearized augmented plane-wave (FLAPW) method \cite{kurz01} or
the projector augmented-wave (PAW) method \cite{hobbs}. Unconstrained non-collinear calculations are
also performed within the atomic sphere approximation (ASA) by using a real-space linearized muffin-tin 
orbital (LMTO) method \cite{robles,bergman1,bergman2} 
and the Korringa-Kohn-Rostoker (KKR) method~\cite{yavorsky}.
Other works~\cite{co7a,co7b,fpsd} rely on ab initio spin-dynamics in terms of a constrained LSDA 
by means of a fully relativistic KKR
method and solving simultaneously the Landau-Lifshitz-Gilbert equations for the evolution
of the orientations of magnetic moments.
Although such simulations are very accurate in finding the magnetic ground state of complex systems, they are
very costly and, in practice, require a massively parallel computer architecture.

Multiscale approaches based on a first principles evaluation of model parameters
are very useful to study both the ground state and the dynamics of spin-systems.
In Refs. \onlinecite{mc-cluster-a,mc-cluster-b,mc-cluster-c} the torque method \cite{torque} was 
employed to calculate isotropic exchange interactions,
and then Monte-Carlo simulations were performed
to study temperature dependent magnetism of nanoclusters.
This approach can, in principle, be extended to include relativistic contributions to the
exchange interactions \cite{rtorque}. Nevertheless, because of the low (or even missing) 
symmetry of nanoparticles the determination of the exchange coupling and on-site anisotropy matrices 
becomes quite complicated. 
Moreover, as found, e.g., for
Mn and Cr monolayers on Cu(111) higher order spin-interactions are needed for
an accurate mapping of the energy obtained from first principles calculations~\cite{kurz01}. 
Recently, a fast ab initio approach that makes use of a suitable parametrization of the 
configurational energy of a complex magnetic system,
namely, a spin cluster expansion (SCE), 
has been proposed \cite{sce1,sce2}, but not yet applied intensively. 

In this work we introduce a new scenario to construct parameters of a spin-model containing
interactions, in principle, up to arbitrary order. Our method is based on relativistic first principles 
calculations of the energy, whereby a sufficiently large number of states with different non-collinear
magnetic configurations (orientational states) are considered to enable a least square fit of
the parameters of the spin-model. In order to determine the magnetic ground state of the system  
we then solve the Landau-Lifshitz-Gilbert equations derived from the corresponding spin Hamiltonian.

The half-filled valence configuration of Cr yields a large magnetic moment and strong
antiferromagnetic inter-atomic bonding leads in turn to magnetic frustration and complex spin phenomena. 
The simplest system exhibiting such properties is a trimer.  
The non-collinear magnetic structure of supported triangular clusters has been first
investigated by means of a self-consistent vector Anderson model \cite{uzdin}. 
First principles calculations of an equilateral Cr trimer supported on a Au(111) surface
have also confirmed a frustrated non-collinear magnetic structure \cite{gotsis,bergman1} 
and revealed a collinear antiferromagnetic magnetic ground state
for a linear chain of three Cr atoms~\cite{bergman1}.

We apply our new method to Cr trimers deposited on a Au(111) surface in equilateral,
linear and isosceles geometries. Though these systems are governed by large antiferromagnetic
nearest neighbor couplings, we intended to trace the role of the relativistic 
interactions in the formation of the magnetic ground state. 
Prominently, for an equilateral trimer the Dzyaloshinsky-Moriya interactions are shown to fix the 
chirality of the magnetic ground state, whereas in cases of 
linear and isosceles trimers the inter- and on-site anisotropic
terms lead to an in-plane orientation of the antiferromagnetic ground state.

%%%%%%%%%%%%%%%%%%%%%%%%%%%%%%%%%%%%%%%%%%%%%%%%%%%
%Theoretical approaches and computational methods %
%%%%%%%%%%%%%%%%%%%%%%%%%%%%%%%%%%%%%%%%%%%%%%%%%%%

\section{Theoretical approaches and computational methods}
\label{sec:theo}

\subsection{The energy of a classical spin-system}

Neglecting intraatomic non-collinearity, the magnetic state of $N$ atoms
is described by the array $\{ \vec{M}_i \}_{i=1,\dots,N}$, where
$\vec{M}_i=M_i \, \vec{\sigma}_i$ ($|\vec{\sigma}_i|=1$) is the magnetic moment of
a particular atom labeled by $i$. In a large class of magnetic systems, referred to as 'good moment'
systems, the longitudinal fluctuations of the moments can also be neglected,
i.e., the magnitudes of the moments, $M_i$, can be considered independent of the 
orientational state, $\{ \vec{\sigma}_i \}_{i=1,\dots,N}$.
The most general expression of the energy up to second
 order of the spin-variables can be written as
\be
 \label{eq:E-02}
E(\{ \vec{\sigma}_i \})= E^{(0)} + E^{(2)}(\{ \vec{\sigma}_i \}) \; ,
\ee
with
\be
 \label{eq:E2}
  E^{(2)}(\{ \vec{\sigma}_i \})=
  \frac{1}{2} \sum_{i \ne j} \vec{\sigma}_i \, {\bf J}_{ij} \, \vec{\sigma}_j
  + \sum_{i} \vec{\sigma}_i \, {\bf K}_{i} \, \vec{\sigma}_i \; ,
\ee
where the ${\bf J}_{ij}=\{J_{ij}^{\alpha\beta}\}$ $(\alpha,\beta=x,y,z)$ 
are the generalized exchange interaction matrices, the
${\bf K}_{i}=\{K_{i}^{\alpha\beta}\}$ are the (second-order) on-site anisotropy
constant matrices.
Within a non-relativistic approach the on-site anisotropy constants vanish, as well as
the exchange tensor takes a simple diagonal form, ${\bf J}_{ij}=J_{ij} \, {\bf I}$ with ${\bf I}$
being the unit matrix, thus, the isotropic Heisenberg model
is acquired. 
For a transparent physical interpretation the exchange tensor, $\mathbf{J}_{ij}$, can be decomposed
into three terms as~\cite{rtorque}
\begin{equation}
\mathbf{J}_{ij} = J_{ij} \mathbf{I} + \mathbf{J}^{S}_{ij} + \mathbf{J}%
^{A}_{ij} \; ,
\end{equation}
where $J_{ij}$ is the isotropic part of the exchange tensor,
\begin{equation}
J_{ij}=\frac{1}{3} Tr\left(  \mathbf{J}_{ij}\right)  \; ,
\end{equation}
the traceless symmetric anisotropic exchange tensor, $\mathbf{J}^{S}_{ij}$  is defined as
\begin{equation}
\mathbf{J}^{S}_{ij}= \frac{1}{2} \left(  \mathbf{J}_{ij} + \mathbf{J}%
_{ij}^{T} \right)  - J_{ij} \mathbf{I} \; ,
\label{eq:JSij}
\end{equation}
where $T$ denotes transpose of a matrix,
while the antisymmetric exchange matrix, $\mathbf{J}^{A}_{ij}$, as
\begin{equation}
\mathbf{J}^{A}_{ij}= \frac{1}{2} \left(  \mathbf{J}_{ij} - \mathbf{J}%
_{ij}^{T} \right)  \; .
\end{equation}
The antisymmetric part of the intersite exchange interaction can then be represented as,
\begin{equation}
\vec{\sigma}_{i} \, \mathbf{J}^{A}_{ij} \, \vec{\sigma}_{j} = 
\vec{D}_{ij} \left(
\vec{\sigma}_{i} \times\vec{\sigma}_{j} \right)  \;,
\label{eq:DMinteraction}
\end{equation}
which is the well-known relativistic Dzyaloshinsky--Moriya (DM) interaction
\cite{Dzyalo-58,Moriya-PR60}, with the vector $\vec{D}_{ij}$ defined as
\begin{equation}
\label{eq:DM-vector}D^{x}_{ij}=\frac{1}{2} \left(  J^{yz}_{ij}-J^{zy}_{ij}
\right)  , \; D^{y}_{ij}=\frac{1}{2} \left(  J^{xz}_{ij}-J^{zx}_{ij} \right)
, \; D^{z}_{ij}=\frac{1}{2} \left(  J^{xy}_{ij}-J^{yx}_{ij} \right)  \; .
\end{equation}
The asymmetric exchange interactions induced by 
the spin-orbit coupling have been shown to
have crucial consequences to the magnetic ground state 
in thin films.\cite{bode,mnw}
For transition metal clusters such effects are expected to be even more important 
due to their reduced symmetry.

Unlike most of the thin films with uniaxial or biaxial symmetry, 
in case of finite clusters the structure of the 
on-site anisotropy matrices can not, in general, be predicted 'a priori', i.e., from symmetry principles.
The on-site anisotropy can, at best, be characterized by diagonalizing the matrix ${\bf K}_i$,
\be
 \label{eq:onsite}
\vec{\sigma}_i \, {\bf K}_i \, \vec{\sigma}_i = \sum_\lambda K_i^\lambda \, 
(\vec{\sigma}_i \cdot \vec{e}^{\: \lambda}_i)^2 \; ,
\ee
where $K_i^\lambda$ and the unit vectors $\vec{e}^{\: \lambda}_i$ ($\lambda=1,2,3$) are the eigenvalues and 
corresponding eigenvectors 
of ${\bf K}_i$. Clearly, the easy axis is associated by the eigenvector that refers to the minimum value
of $K_i^\lambda$. Note that the matrix ${\bf K}_i$ can be chosen to be symmetric, therefore, the
eigenvectors $\vec{e}^{\: \lambda}_i$ are pairwise normal to each other.  
Obviously, the symmetric anisotropic exchange interaction, see  Eq.~(\ref{eq:JSij}),  can be decomposed 
in a similar way,
\be
 \label{eq:symmx}
\vec{\sigma}_i \, {\bf J}^S_{ij} \, \vec{\sigma}_j = \sum_\lambda J_{ij}^{S,\lambda} \, 
(\vec{\sigma}_i \cdot \vec{e}^{\: \lambda}_{ij}) 
(\vec{\sigma}_j \cdot \vec{e}^{\: \lambda}_{ij}) \; ,
\ee
with $J_{ij}^{S,\lambda}$ and $\vec{e}^{\: \lambda}_{ij}$ being the eigenvalues and eigenvectors
of the matrix ${\bf J}^S_{ij}$, respectively.

The second order approximation, Eq.~(\ref{eq:E2}), 
is, however, not always sufficient to describe the energy of a magnetic system.\cite{kurz01}
As will be shown  
adding a term, $E^{(4)}$, to Eq.~(\ref{eq:E-02}) corresponding to the fourth order spin-interactions 
considerably improves the quality of the mapping
of the energy from first principles calculations to the spin-model. In order to 
keep our model tractable, we extended Eq.~(\ref{eq:E-02}) only by SU(2)
invariant fourth order terms,
%\be
% \label{eq:E-024}
%E(\{ \vec{\sigma}_i \})= E^{(0)} + E^{(2)}(\{ \vec{\sigma}_i \}) + 
%E^{(4)}(\{ \vec{\sigma}_i \}) \; ,
%\ee
\be
 \label{eq:E4}
E^{(4)}(\{ \vec{\sigma}_i \}) =
         \sum_{\begin{subarray}{c} i,j,k,l \\ (i < j, k < l) \end{subarray}} 
Q_{ijkl}(\vec{\sigma}_i \cdot \vec{\sigma}_j)
        (\vec{\sigma}_k\cdot \vec{\sigma}_l) \; .
\ee
It is easy to see that in case of three atoms the above sum consists of only six different terms,
 therefore, the following simplified notation can be used, 
\[ 
Q_{12}=Q_{1212}, \: Q_{13}=Q_{1313},\:  Q_{23}=Q_{2323}, \: 
Q^1_{23}=Q_{1213}, \: Q^2_{13}=Q_{2123}, \: Q^3_{12}=Q_{3132} \; .
\]
We determined the parameters, $J_{ij}^{\alpha \beta}$, $K_i^{\alpha \beta}$, $Q_{ij}$ and
$Q^{i}_{jk}$ for different Cr trimers on a Au(111) surface
by fitting the energy of the orientational states obtained from first principles calculations
to Eq.~(\ref{eq:E2}) supplemented by the terms,  Eq.~(\ref{eq:E4}).

\subsection{Evaluation of the parameters for Cr trimers}
\label{sec:KKREC}

\begin{figure}[ht!]
\begin{tabular}{lcr}
\includegraphics[width=5cm,clip]{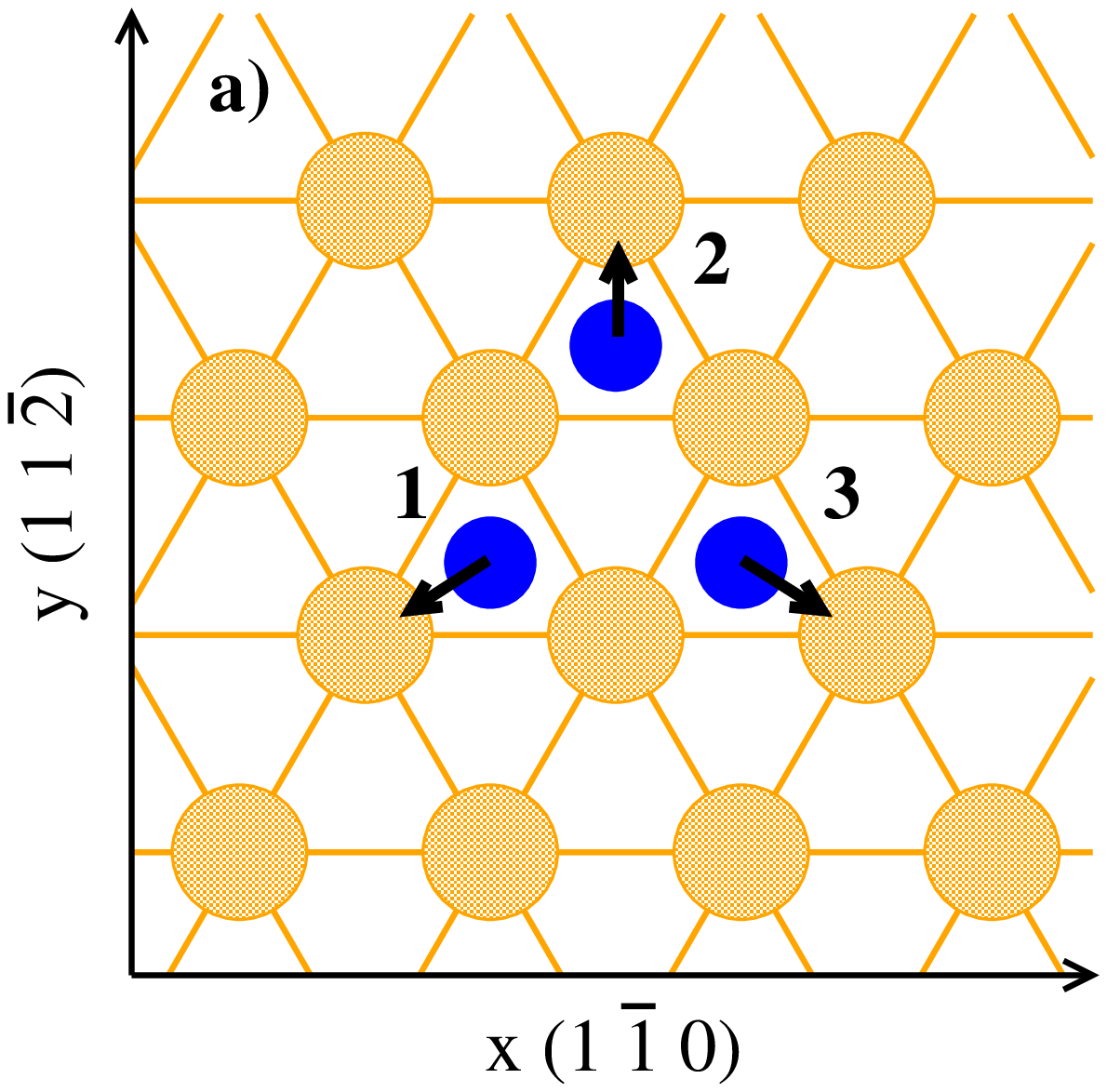} &
\includegraphics[width=5cm,clip]{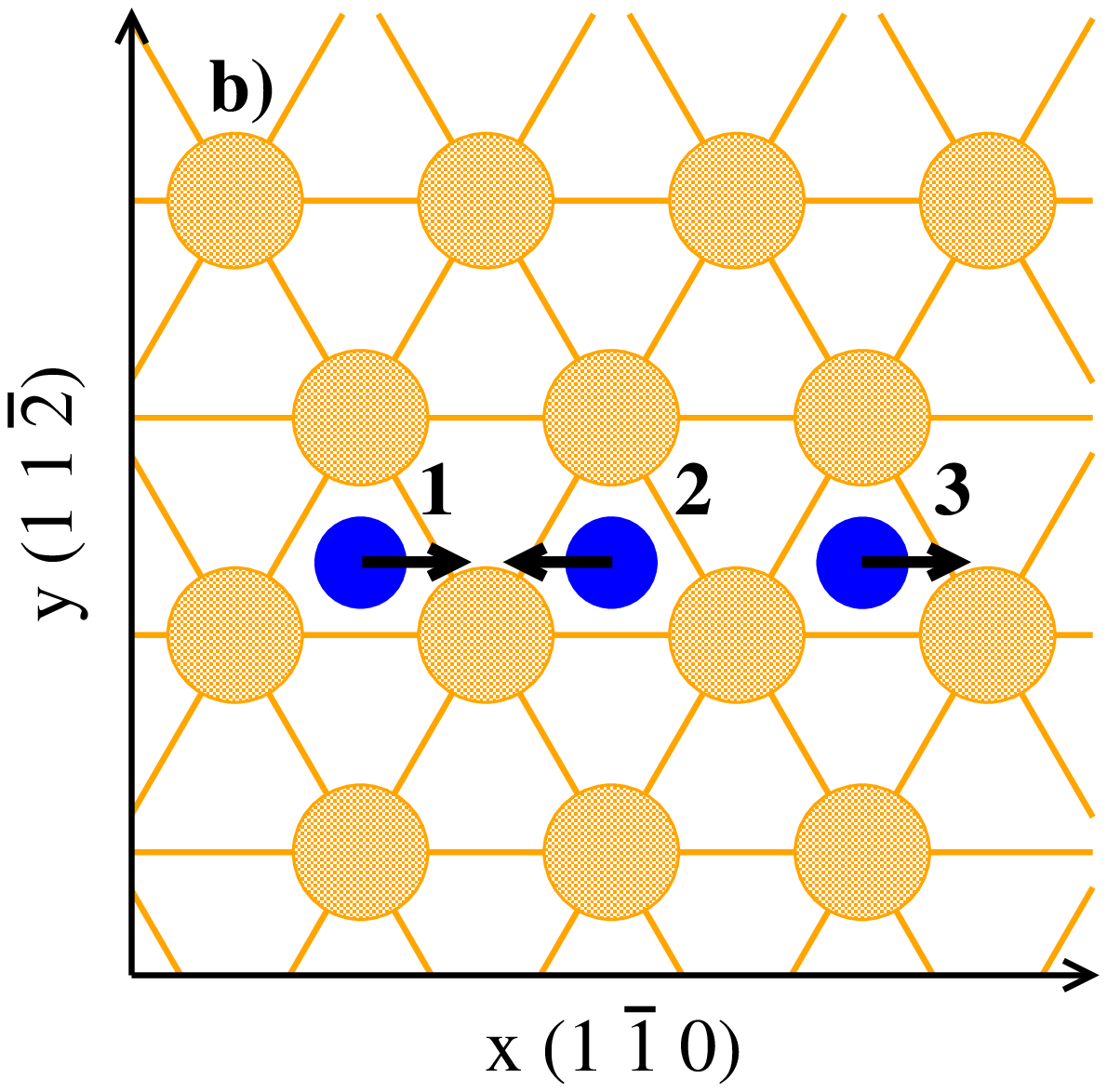} &
\includegraphics[width=5cm,clip]{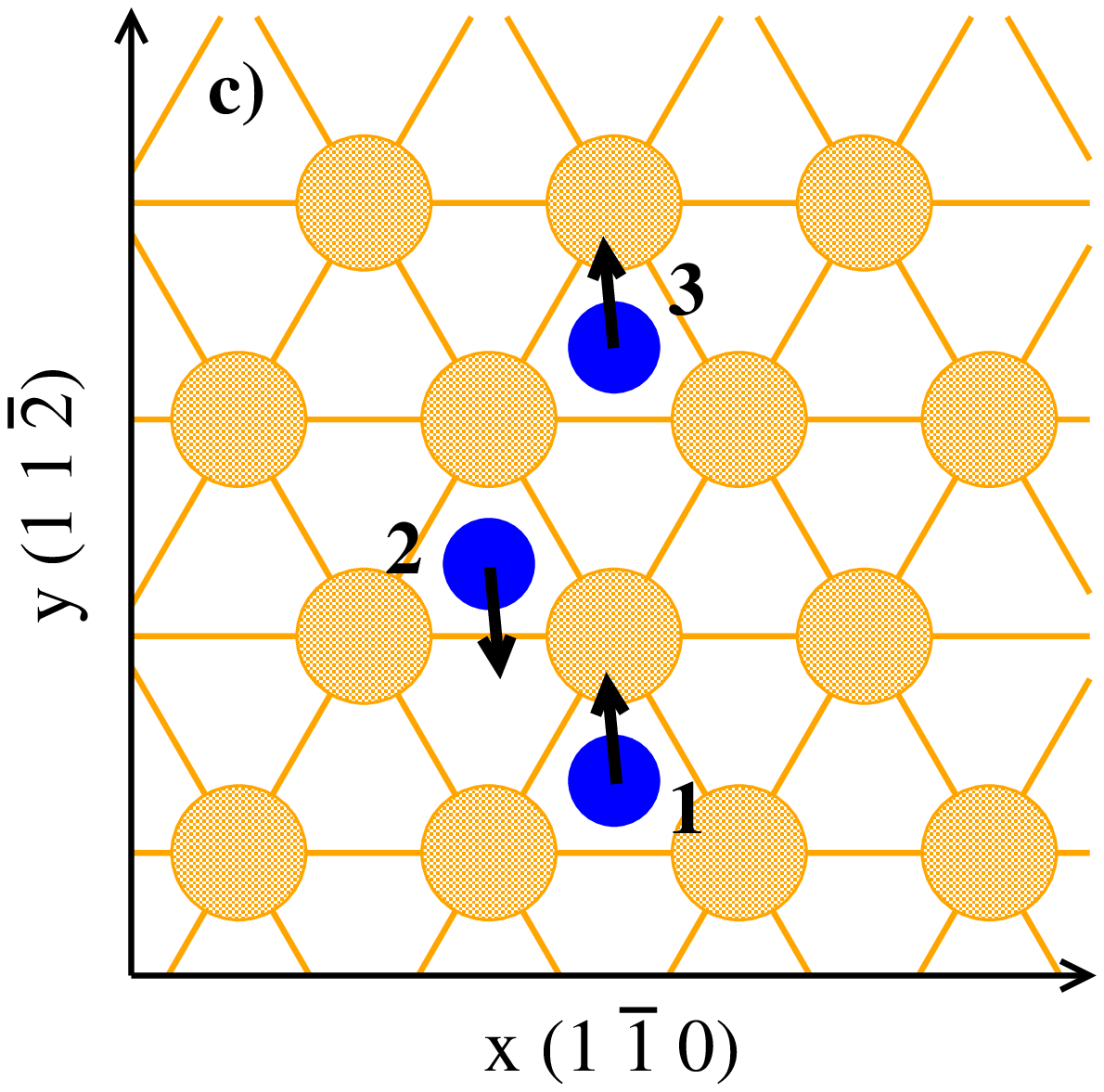}
\end{tabular}
\vskip -0.3cm
\caption{(Color online)
Geometry of the Cr trimers (numbered solid circles) deposited on top of a Au(111) surface layer
(patterned circles) considered in the present work: {\bf a)} equilateral triangle,
{\bf b)} linear chain, and {\bf c)} isosceles triangle.
The arrows denote the ground state orientation of the magnetic moments of the Cr atoms. 
   \label{fig:clusters}}
\end{figure}

%Embedded cluster method
In order to calculate the electronic structure of the Cr$_3$ clusters 
we applied an Embedded Cluster Green's function technique as combined
with the Korringa-Kohn-Rostoker method (KKR-EC) \cite{LSW02}.
Within the KKR-EC the matrix of the so--called scattering path operator (SPO),
$\mbox{\boldmath $\tau$}_{\mathcal{C}}$, corresponding to a finite
cluster $\mathcal{C}$ embedded into a host system can be obtained from
the following Dyson equation,
\begin{equation}
\mbox{\boldmath $\tau$}_{\mathcal{C}}(E)=\mbox{\boldmath
$\tau$}_{h}(E)\left[
\mbox{\boldmath $I$}-(\mbox{\boldmath $t$}_{h}^{-1}%
(E)-\mbox{\boldmath $t$}_{\mathcal{C}}^{-1}(E))\mbox{\boldmath
$\tau$}_{h}%
(E)\right]  ^{-1}\quad, \label{eq:dyson}%
\end{equation}
where $\mbox{\boldmath $t$}_{h}(E)$ and $\mbox{\boldmath $\tau$}_{h}(E)$
denote the single--site scattering matrix and the SPO matrix for the
pristine host confined to the sites in $\mathcal{C}$, respectively,
while $\mbox{\boldmath $t$}_{\mathcal{C}}$ comprises the single--site
scattering matrices of the embedded atoms. 
Note that Eq.~(\ref{eq:dyson}) accounts for all
scattering events in the system merging the cluster and the host.
Once $\mbox{\boldmath $\tau$}_{\mathcal{C}}$ is derived, all quantities
of interest for a cluster, i.e. the charge and magnetization densities,
the spin-- and orbital moments as well as the exchange interaction energy 
can be calculated.
The electronic structure of the host gold surface including three layers
of empty spheres to represent the vacuum region was calculated in terms of
the fully relativistic screened Korringa-Kohn-Rostoker method
\cite{SUW+94a,SUW+94b}.
The cluster calculations were then carried out such that the 
Cr atoms substituted empty spheres 
on top of the surface, whereas
no attempts were made to include lattice relaxation effects. 
In Fig.~\ref{fig:clusters} shown is the geometry of the three Cr trimers considered in the
present work, namely, forming an equilateral triangle, a linear chain and an
isosceles triangle.

The local spin--density approximation as parametrized by Vosko
\emph{et al. }\cite{VWN80} was applied, the effective
potentials and fields were treated within the atomic sphere
approximation. When solving the Kohn-Sham-Dirac equation and also for
the multipole expansion of the charge densities we used a cut--off of
$\ell_{max}=2$. When performing self-consistent calculations for the linear chain and 
the isosceles triangle we fixed 
the direction of the magnetization on all the three Cr atom normal to the surface,
whilst for the equilateral trimer we used the 120$^\circ$ N\'eel
state indicated by arrows in Fig.~\ref{fig:clusters}a as reference (see Sec. III.A).

%As we checked in several cases the magnitude of the magnetic moments were proved
%fairly independent on the magnetic configurations.
For the calculation of the energy of the orientational states we applied
the magnetic force theorem \cite{DKS90,SUW95,Jan99} by using the self--consistent
potentials determined in the reference state. In here, only band-energy differences have
to be calculated requiring, however, a high precision for the necessary
Brillouin zone integrals.\cite{LSW02} 
To this end, when evaluating $\mbox{\boldmath $\tau$}_{h}(E)$
we used over 3300 $k$-points in the irreducible (1/6) segment of the Surface Brillouin zone.

In order to determine the parameters of our spin-model, we generated 
a large number of random magnetic configurations, $\{\vec{\sigma}_i^n\},\quad n=1,\dots,N$, 
and calculated the corresponding band-energies~\cite{LSW02},
$E_b^n=E_b(\vec{\sigma}_1^n,\vec{\sigma}_2^n,\vec{\sigma}_3^n)$.
Introducing a (row--) vector containing all combinations of the components $\vec{\sigma}_{i,\alpha}^n$ 
$(\alpha=x,y,z)$ occurring in  expressions Eqs.~(\ref{eq:E2}) and (\ref{eq:E4}),
\begin{eqnarray}
 \mathbf{X}_n &=& (
    \sigma_{1,x}^n\sigma_{2,x}^n,
    \sigma_{1,x}^n\sigma_{2,y}^n,
    \sigma_{1,x}^n\sigma_{2,z}^n,
    \dots
    \sigma_{1,x}^n\sigma_{1,x}^n, \nonumber \\
    && \dots,\sum_{i,j=x,y,z} \sigma_{3,i}^n\sigma_{1,i}^n\sigma_{3,j}^n\sigma_{2,j}^n ) \; ,
\end{eqnarray}
and a vector of the corresponding parameters of the model,
\begin{equation}
 \mathbf{P} = \left ( J^{12}_{xx},J^{12}_{xy},J^{12}_{xz},\dots, K^{1}_{xx},\dots, 
Q^{3}_{12} \right ) \; ,
\end{equation} 
the energy of the $nth$ configuration can simply be written as 
\be \label{eq:En}
E_{n} =  \mathbf{P} \, \mathbf{X}_n^T  \; .
\ee
An optimal choice of the parameters should minimize the difference (error) between the calculated
band-energies, $E_n^b$, and the energy related to the spin-model, $E_n$. The square of error is defined as
\begin{equation}
\Delta E^2 = \sum_{n=1}^N \left (E^n - E^b_n \right )^2 \; ,
\end{equation}
and by substituting Eq.~(\ref{eq:En}) a least square condition leads to the solution,
\begin{equation}
 \mathbf{P} = 
 \sum_n E^b_n \, \mathbf{X}_n \, 
\left ( \sum_n  \mathbf{X}_n^{T} \, \mathbf{X}_{n} \right )^{-1}
\; .
\end{equation} 
The number of considered random configurations can be
increased until the parameters achieve well converged values.

\begin{figure}[ht!]
\includegraphics[width=8cm,clip]{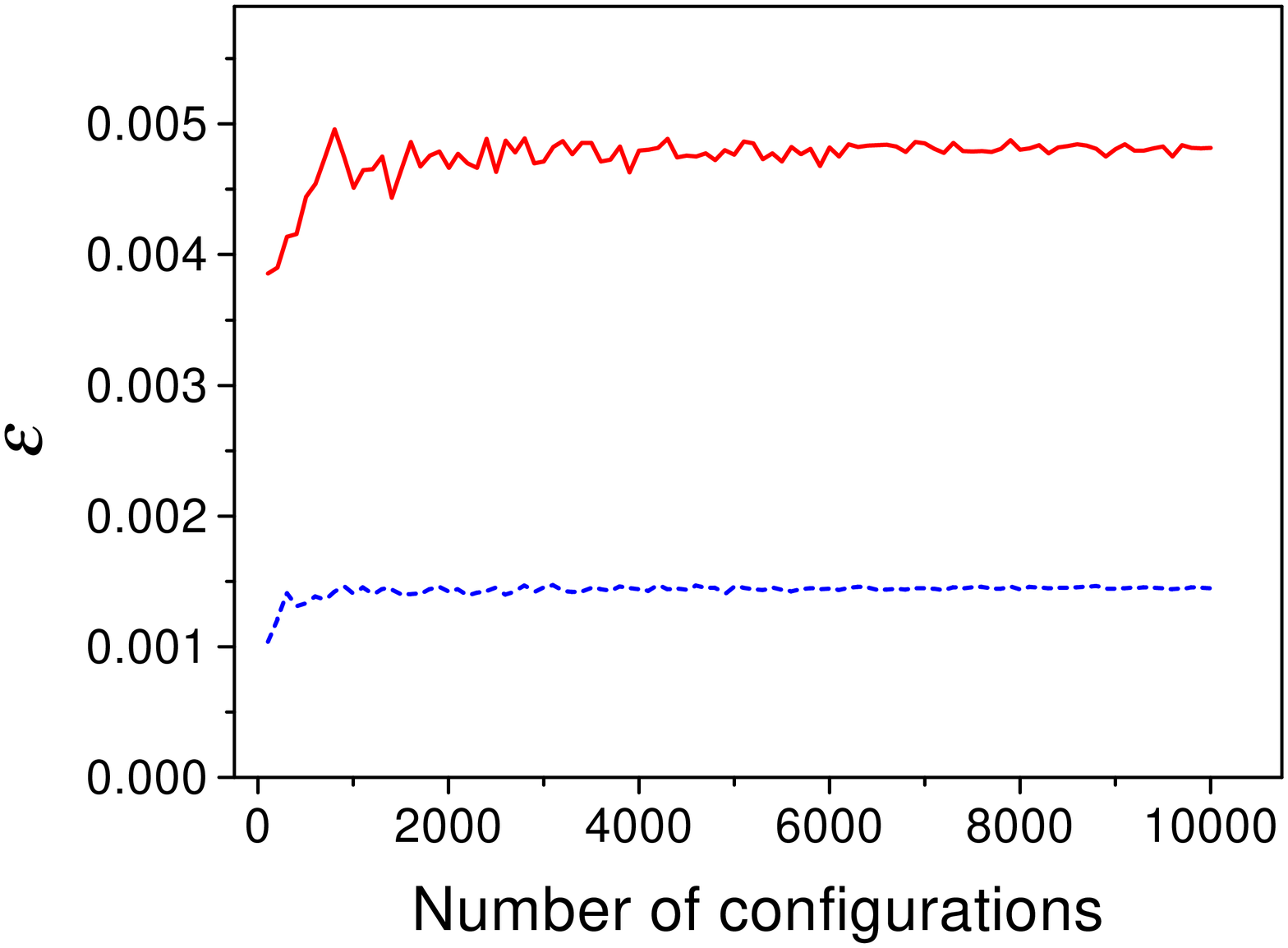}
\vskip -0.3cm
\caption{(Color online) 
The evolution of the relative error, $\varepsilon = \sqrt{\Delta E^2/\langle E^2 \rangle}$, of the fitting procedure 
against the number of considered configurations for the isosceles Cr trimer.
Solid (red) and dashed (blue) lines refer to the parameter set
excluding and including SU(2) invariant bi-quadratic terms in the spin-model, respectively.
}
\label{fig:error} 
\end{figure}
The quality of the fit is characterized by the relative error,
$\varepsilon = \sqrt{\Delta E^2/\langle E^2 \rangle}$,
where $\langle E^2 \rangle$ is the average of $E_n^2$ over all the configurations. 
This error as a function of the number of configurations, $N$, looks very similar for all the 
three clusters. 
As it is shown in Fig.~\ref{fig:error}  for a Cr trimer forming an isosceles triangle the error of 
the fit is well stabilized around 0.48~\% above $N \simeq 5000$ when only the second order spin-interactions
are taken into account, see Eq.~(\ref{eq:E2}).  The error, however, reduced
to 0.14~\% when also the SU(2) invariant fourth order terms, Eq.~(\ref{eq:E4}),
are considered and only about 2000 configurations were sufficient  
to get a stable error.

\begin{figure}[ht!]
\includegraphics[width=8cm,clip]{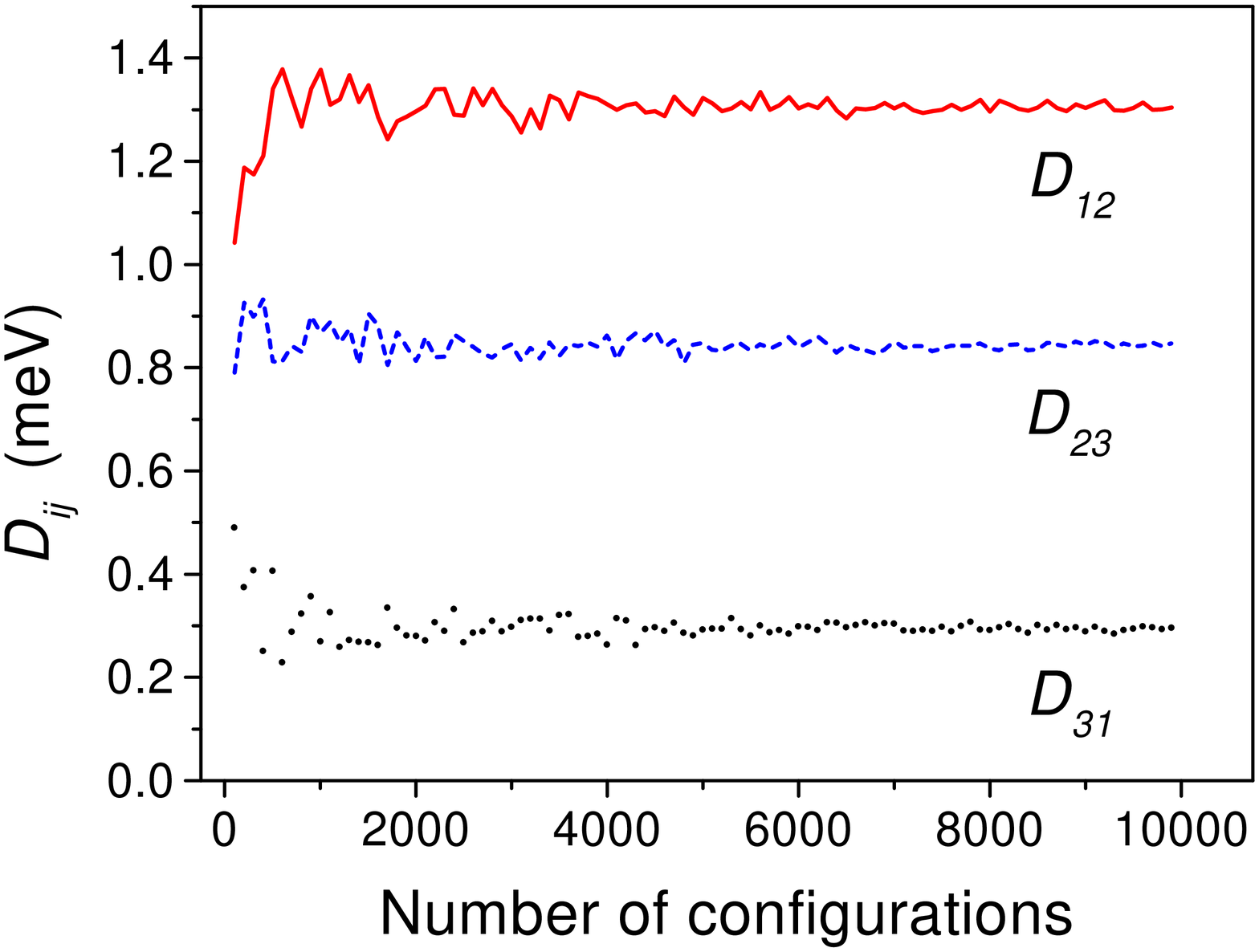}
\vskip -0.3cm
\caption{(Color online) Convergence of the magnitudes of the
Dzyaloshinsky-Moriya vectors, $D_{ij}$, against the number of configurations for the case of an
isosceles Cr trimer. The labels of the DM vectors refer to the Cr atoms as numbered in 
Fig.~\ref{fig:clusters}c.
}
\label{fig:DM_conv}
\end{figure}
The configurational energy of the Cr trimers is dominated by quite large antiferromagnetic 
isotropic exchange interactions, $J_{ij}\simeq 100-150$ meV. In general, we obtained 
by two order smaller DM interactions, $D_{ij} \simeq 0.5-2.0$ meV, whereas the typical
range of the anisotropic symmetric exchange interactions
and the on-site anisotropy constants was about 0.1 meV or even less.
This trend can be understood in terms of a perturbation treatment with respect to the 
spin-orbit coupling parameter, $\xi$, since the DM interactions turn to be proportional with $\xi$,
whereas the anisotropy terms appear (at best) in second order of $\xi$.
Obviously, a required relative accuracy for the small interaction parameters
can be achieved at a much larger number of configurations than 
for the total configurational energy.
This is demonstrated in Fig.~\ref{fig:DM_conv} showing the evolution of the DM interactions 
for an isosceles Cr trimer. As can be inferred from this figure,
about 6-7000 configurations are needed to stabilize the values of $D_{ij}$ with a
relative accuracy of 1~\%.
In order to reach the same relative accuracy 
for the coefficients with the smallest magnitude, namely, 
for the in-plane on-site anisotropy constants,
we had to generate about 10000 random configurations.

\subsection{Determination of the magnetic ground state}

Once the parameters of the spin model are fixed,
the ground state configuration of the system can easily be determined by solving
the Landau-Lifshitz-Gilbert (LLG) equations for the transversal components of 
the magnetizations,
\begin{equation}
\frac{\partial\vec{\sigma}_i}{\partial t}=
-\frac{\gamma}{1+\alpha^2} \, \vec{\sigma}_i \times \vec{H}_i^{eff} -
\frac{\alpha \gamma}{1+\alpha^2} \, \vec{\sigma}_i \times 
\left( \vec{\sigma}_i \times \vec{H}_i^{eff} \right) \; ,
\label{eq:LLG}
\end{equation}
where $\gamma$ and $\alpha$ are the gyromagnetic ratio and the Gilbert damping factor, respectively,
whereas the effective fields, $\vec{H}_i^{eff}$, are defined as
\be
\vec{H}_i^{eff} = -\frac{1}{M_i} \frac{\partial E \left( \left\{ \vec{\sigma} \right\} \right) }
{\partial \vec{\sigma}_i} \; ,
\label{eq:Heff}
\ee
by using Eqs.~(\ref{eq:E2}) and (\ref{eq:E4}). % and (\ref{eq:E-024}).
%Clearly, in the ground-state or in a metastable state of the system, $\vec{H}_i^{eff}=0$, which
%implies $\partial\vec{\sigma}_i/\partial t = 0$.
%We found this condition by solving Eq.~(\ref{eq:LLG}) via a fourth-order Runge-Kutta method. %It should be noted that the phenomenological parameters,
%$\gamma$ and $\alpha$, were set to optimize the stability and the speed of the numerical procedure
%to obtain the ground-state.
It should be stressed that in the present context Eq.~(\ref{eq:LLG}) is merely used as a numerical tool 
to find the energy minimum in the six-dimensional phase-space, $\{ \vec{\sigma}_i \}$, describing the 
non-collinear configurations of the Cr trimers. 
The stability and the speed of the applied numerical procedure, a fourth-order 
Runge-Kutta method, for integrating Eq.~(\ref{eq:LLG}) was, therefore, optimized by adjusting 
the phenomenological parameters, $\gamma$ and $\alpha$.

%%%%%%%%%%%%%%%%%%%%%%%%%%%%
%RESULTS AND DISCUSSION  %
%%%%%%%%%%%%%%%%%%%%%%%%%%%%
\section{Results and Discussions}
\label{sec:results}

\subsection{Equilateral trimer}

We first investigated a Cr trimer forming an equilateral triangle on top of Au(111) as 
shown in Fig.~\ref{fig:clusters}a. 
Since our previous first principles spin-dynamics calculations~\cite{fpsd}
resulted in a 120$^\circ$ N\'eel type of ground state, see Fig.~\ref{fig:clusters}a,
in here we used this configuration as a reference state to determine effective potentials and
fields self-consistently. Reassuringly, however,
the calculated spin magnetic moments of the Cr atoms,    
4.4~$\mu_B$, were proved to be practically independent 
from the magnetic configuration of the trimer.
Since in this case both the substrate and the trimer exhibit a $c_{3v}$ 
point group symmetry, the exchange interaction matrices, ${\bf J}_{12}$, ${\bf J}_{13}$ and
${\bf J}_{23}$, as well as the on-site anisotropy matrices, ${\bf K}_{1}$, ${\bf K}_{2}$ and
${\bf K}_{3}$, are linked in terms of appropriate similarity transformations. 
The number of independent parameters of the model is, therefore,
considerably reduced, e.g., the isotropic exchange parameters become identical 
and the on-site anisotropy matrix 
corresponding to the atom labeled by 2 in Fig.~\ref{fig:clusters}a is of the form,
\begin{equation}
{\bf K}_2=
\left[\begin{array}{ccc}
K_{xx}& 0& 0\\
0&K_{yy}&K_{yz}\\
0&K_{yz}&K_{zz} \end{array} \right] \; .
\label{eq:K2}
\end{equation}
As a consistency check of our fitting process, 
the obtained parameters satisfied  all the relationships dictated by the symmetry of the system.

\begin{figure}[ht!]
\includegraphics[width=6cm,bb=10 10 285 240,clip]{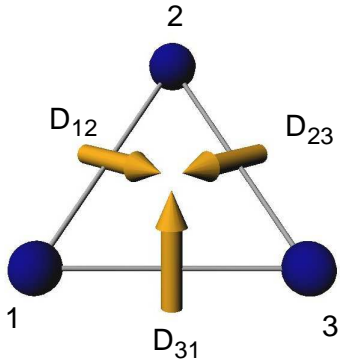}
\vskip -0.3cm
 \caption{(Color online) Schematic view of the DM vectors for an equilateral Cr trimer.
\label{fig:dm-eq}}
\end{figure}

The dominant parameters determining the ground state of this Cr trimer are
the isotropic exchange interactions, $J_{12}=J_{13}=J_{23}=144.9$ meV, and
the DM interactions with magnitudes, $D_{12}=D_{13}=D_{23}=1.78$ meV.
As mentioned earlier, the on-site anisotropy terms are much smaller in magnitude, e.g.
$K_{xx}=-0.09$ meV in Eq.~(\ref{eq:K2}). The DM vectors, visualized in Fig.~\ref{fig:dm-eq},
point towards a common point lying above the geometrical center of the Cr trimer.
This is the consequence of Moriya's second rule for the DM vectors \cite{Moriya-PR60}: 
if a mirror plane of the system is bisecting the line between a pair of sites
then the respective DM vector lies in the mirror plane.
Noteworthy, the coefficients of the biquadratic 
spin-interactions are as follows: $Q_{12}=Q_{13}=Q_{23}=-4.42$ meV and 
$Q^1_{23}=Q^2_{13}=Q^3_{12}=7.06$ meV.

By solving the LLG equations as described in Section II.C with the above
parameters we indeed obtained the ground state indicated in Fig.~\ref{fig:clusters}a, namely, a
state which is very close to an in-plane 120$^\circ$ N\'eel state with almost negligible
out-of-plane components of the magnetic moments.
Quite obviously, an equivalent ground state can be generated from this state by 
simultaneously reversing the directions of all the magnetic moments.      

\begin{figure}[ht!]
\begin{tabular}{ccc}
\includegraphics[width=4cm,clip]{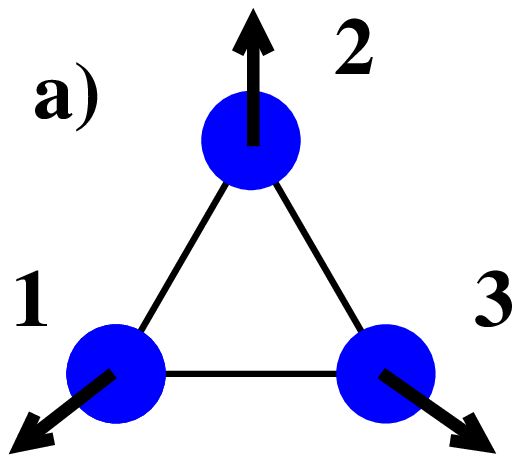} & \ \hskip 1.0cm \ &
\includegraphics[width=4cm,clip]{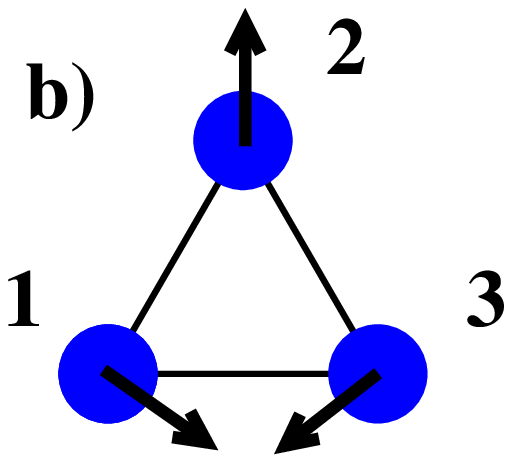} 
\end{tabular}
%\vskip -0.3cm
 \caption{(Color online) Two typical ground state configurations 
of an equilateral Cr trimer in absence of DM interactions.
The two configurations refer to different chiralities: {\bf a)} $\kappa_z=-1$ and
{\bf b)} $\kappa_z=1$, see Eq.~(\ref{eq:chirality}). 
\label{fig:chirality}}
\end{figure}

Considering only isotropic exchange interactions,
a frustration induced by the geometry of the Cr trimer leads
to an eight-fold degenerate non-collinear ground state.
These states can be divided into two classes as indicated in Fig.~\ref{fig:chirality}.
One class consists of the configuration depicted in Fig.~\ref{fig:chirality}a
and the corresponding one with reversed directions (two configurations), the second
class consists of that in Fig.~\ref{fig:chirality}b and those generated from this state
via $c_{3v}$ symmetry transformations and time reversal (six configurations in total).
Defining the chirality vector of the system as 
\begin{equation}
\vec{\kappa} = \frac{2}{3\sqrt{3}} \sum_{(ij)}
          (\vec{\sigma}_i\times\vec{\sigma}_j) \; ,
\label{eq:chirality}
\end{equation}
where the summation runs over the three directed bonds, (12, 23 and 31), forming the triangle,
the two classes can be assigned to chiralities $\kappa_z=-1$ and $\kappa_z=1$,
respectively. 
Note that for in-plane configurations the vector $\vec{\kappa}$ is normal to the plane of the triangle. 

Recalling Eq.~(\ref{eq:DMinteraction}), when switching on the DM interactions 
the degeneracy of the ground state will evidently be lifted according to the chirality, $\kappa_z$.
In the present case of in-plane magnetization 
the contribution of the DM interaction to the energy can simply be expressed as 
\begin{equation} \label{eq:edm}
E_{DM} = \frac{3\sqrt{3}}{2}\, D_z \, \kappa_z \; ,
\end{equation}
where $D_z$ denotes the $z$ component of any DM vector.
Since according to our calculations $D_z=0.97$ meV, 
the states with $\kappa_z=-1$
depicted in Fig.~\ref{fig:chirality}a become the ground state of the system,
while the states with $\kappa_z=1$, see Fig.~\ref{fig:chirality}b, are higher
in energy by $\Delta E=5.04$ meV. Thus the ground state we found by solving the LLG equations
is a simultaneous consequence of antiferromagnetic exchange interactions and the DM interactions 
as shown in Fig.~\ref{fig:dm-eq}. 

Finally in this section, a note has to be added concerning the reference state for the fitting
procedure described in Sec.~II.B. As discussed in quite some details in Ref.~\onlinecite{fpsd}
when choosing a normal-to-plane ferromagnetic reference state an erroneous ground state,
Fig.~\ref{fig:chirality}b, was obtained. The very reason of this result is that in this case
the orientations of the fitted DM vectors differ from those depicted in Fig.~\ref{fig:dm-eq},
namely, yielding $D_z<0$. Clearly from Eq.~(\ref{eq:edm}), the energy of the states
related to $\kappa_z=1$ are lowered with respect to those related to $\kappa_z=-1$.
This observation clearly indicates that, in particular, for systems with metastable states 
close to the ground state
one has to be very careful when choosing the reference state serving as basis for subsequent
magnetic force theorem calculations.

\subsection{Linear trimer}

A linear chain of three Cr atoms on top of the Au(111) surface has been considered 
in the geometry as shown in Fig.~\ref{fig:clusters}b. 
As can be seen from this figure this system has only a 
mirror plane normal to the surface and bisecting the chain. 
The calculated magnetic moments are only slightly different from those for the equilateral
triangle: 4.45 $\mu_B$ at the edges of the chain and 4.47 $\mu_B$ at the central atom.
Quite clearly from Table~\ref{tbl:par-lin}, the mirror symmetry imposes
some relationship between the parameters, e.g., $J_{12}$=$J_{23}$ or ${\bf J}_{13}$ has only
diagonal elements etc. Similar to the equilateral trimer  there are large antiferromagnetic
isotropic exchange interactions between the nearest neighbors, whilst
the edge atoms are coupled ferromagnetically.
The magnitudes of the nearest neighbor exchange interactions are very similar to those obtained in terms
of a real space LMTO method for Cr dimers~\cite{bergman1}.
Noticeably, also in this case quite large biquadratic terms of type $Q^k_{ij}$ 
were needed to obtain a sufficiently good fit of the band-energy. 

\begin{table}[ht!]
\begin{center}
\begin{tabular}{|r|rrr|r|rrr|}   \hline
\multicolumn{1}{|c}{ $J_{12}$}  & \multicolumn{3}{|c}{${\bf J}^S_{12}$ } & 
\multicolumn{1}{|c|}{ $\vec{D}_{21}$ } & \multicolumn{3}{c|}{${\bf K}_{1}$}  \\ \hline
      &  0.157 & -0.006 & -0.014 &  0.056 & -0.092 &  -0.007 & -0.128         \\
99.59 & -0.006 &  0.005 &  0.063 &  0.487& -0.007 &  -0.141 &  0.013         \\
      & -0.014 &  0.063 & -0.162 &\ -0.578 & -0.128 &   0.013 &  0.233         \\ \hline\hline
\multicolumn{1}{|c}{ $J_{13}$}  & \multicolumn{3}{|c}{${\bf J}^S_{13}$ } & 
\multicolumn{1}{|c|}{ $\vec{D}_{31}$ } & \multicolumn{3}{c|}{${\bf K}_{2}$}  \\ \hline
      &  0.022 &  0.000 &  0.000 &  0.000 &  0.018 &   0.000 &  0.000         \\
-17.85&  0.000 &  0.009 &  0.000 & -0.262 &  0.000 &  -0.066 &  0.034         \\
      &  0.000 &  0.000 & -0.031 &  0.068 &  0.000 &   0.034 &  0.048         \\ \hline \hline
\multicolumn{1}{|c}{ $J_{23}$}  & \multicolumn{3}{|c}{${\bf J}^S_{23}$ } & 
\multicolumn{1}{|c|}{ $\vec{D}_{23}$ } & \multicolumn{3}{c|}{${\bf K}_{3}$}  \\ \hline
      &  0.157 &  0.006 &  0.014 &  0.056 & -0.092 &   0.007 &  0.128         \\
99.59 &  0.006 &  0.005 &  0.063 & -0.487 &  0.007 &  -0.141 &  0.013         \\
      &  0.014 &  0.063 & -0.162 &  0.578 &  0.128 &   0.013 &  0.233         \\ \hline 
\end{tabular}

\medskip

\begin{tabular}{|c|c|c|c|c|c|}   \hline
 \ \quad $Q_{12}$ \ \quad & \ \quad $Q_{23}$ \ \quad & \  \quad$Q_{13}$ \ \quad & \ \quad$Q^1_{23}$ \ \quad &
 \ \quad $Q^2_{13}$ \ \quad & \ \quad $Q^3_{12}$ \ \ \quad  \\ \hline 
 -0.078  & -0.078   & 0.063    & -2.449     &  5.810     & -2.449      \\ \hline
\end{tabular}
\end{center}
\caption{
Calculated isotropic exchange coupling parameters, $J_{ij}$, symmetric anisotropic
exchange tensors, ${\bf J}^S_{ij}$,  DM vectors, $\vec{D}_{ij}$, on-site anisotropy
matrices, ${\bf K}_i$, and SU(2) invariant biquadratic coupling parameters, $Q_{ij}$ and $Q^k_{ij}$,
for a linear Cr trimer. All data are given in units of meV.
\label{tbl:par-lin}
}
\end{table}
\begin{figure}[ht!]
\includegraphics[width=6cm,bb=5 65 275 200,clip]{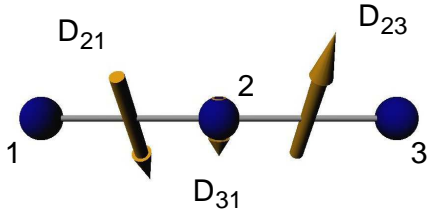}
\vskip -0.3cm
\caption{\label{fig:DM-lin} (Color online) Schematic view of the DM vectors for the linear Cr trimer.}
\end{figure}

Although interactions of relativistic origin have quite minor contributions to the energy,
it is instructive to discuss them in some detail. 
The DM vectors shown schematically in Fig. \ref{fig:DM-lin} are subjects to symmetry conditions:
$\vec{D}_{13}$ lies in the mirror plane, while  $\vec{D}_{21}$ and
$\vec{D}_{23}$ are mirror images of each other as axial vectors. As in the case of the
equilateral triangle, the on-site anisotropy terms and the symmetric anisotropic exchange interactions
have the smallest contributions to the energy. The
on-site anisotropy matrix related to site no.~2 located in the mirror plane has the
structure in Eq.~(\ref{eq:K2}), whereas there is no regularity 
in the matrixelements of ${\bf K}_1$ and ${\bf K}_3$,
except they are related to each other via reflection: $x^\prime=-x$, $y^\prime=y$, $z^\prime=z$.

As expected just by considering isotropic exchange interactions, 
the solution of the LLG equation with sufficiently large damping led to an antiferromagnetic ground state,
 see Fig.~\ref{fig:clusters}b. The direction of the magnetic moments are  parallel 
to the ($1\overline{1}0$) axes which is consistent with symmetry considerations, 
namely, the easy axis of
a collinear magnetic system with a mirror plane should lie either parallel or normal to the
mirror plane~\cite{co7b}. It is important to note that the ambiguity of the reference state mentioned 
in the previous section does not effect the ground state of the linear chain since 
 the DM interactions evidently vanish in a collinear magnetic state.

\subsection{Isosceles trimer}

The third type of Cr trimer we considered is an isosceles triangle as depicted
in Fig.~\ref{fig:clusters}c. 
Apparently, the trimer has a single mirror plane which, however, doesn't 
coincide with the those of the surface layer. 
Therefore the system has no point-group symmetry in this case.
Our calculations resulted in magnetic moments of 4.45~$\mu_B$ for the Cr atoms no. 1 and 3 
and 4.46~$\mu_B$ for the Cr atom no.~2.
As can be inferred from Table~\ref{tbl:par-iso}  
the nearest neighbor isotropic exchange interactions are almost symmetric,
$J_{12} \simeq J_{23}$, and,  as for the linear chain, the second nearest neighbor
isotropic exchange interaction is weakly ferromagnetic. 
The above data indicate that, similar to many transition metal systems, the formation of local moments 
depends mainly on the nearest neighbor environment of the atoms rather than on
long-range interactions or the global symmetry of the system.

\begin{table}[ht!]
\begin{center}
\begin{tabular}{|r|rrr|r|rrr|}   \hline
\multicolumn{1}{|c}{ $J_{12}$}  & \multicolumn{3}{|c}{${\bf J}^S_{12}$ } & 
\multicolumn{1}{|c|}{ $\vec{D}_{12}$ } & \multicolumn{3}{c|}{${\bf K}_{1}$}  \\ \hline
       & -0.126 &  0.058 &  0.047 &\ -0.238 & -0.091 & -0.004 & -0.009   \\
117.97 &  0.058 & -0.035 & -0.022 & -0.472 & -0.004 & -0.042 & -0.029   \\
       &  0.047 & -0.022 &  0.161 &  0.656 & -0.009 & -0.029 &  0.133   \\ \hline\hline
\multicolumn{1}{|c}{ $J_{13}$}  & \multicolumn{3}{|c}{${\bf J}^S_{13}$ } & 
\multicolumn{1}{|c|}{ $\vec{D}_{31}$ } & \multicolumn{3}{c|}{${\bf K}_{2}$}  \\ \hline
       & -0.019 & -0.015 &  0.019 &  0.075 & -0.120 & -0.023 & -0.018   \\
-5.60  & -0.015 & -0.089 & -0.025 &  0.150 & -0.023 & -0.062 & -0.006   \\
       &  0.019 & -0.025 &  0.109 &  0.242 & -0.018 & -0.006 &  0.181   \\ \hline \hline
\multicolumn{1}{|c}{ $J_{23}$}  & \multicolumn{3}{|c}{${\bf J}^S_{23}$ } & 
\multicolumn{1}{|c|}{ $\vec{D}_{23}$ } & \multicolumn{3}{c|}{${\bf K}_{3}$}  \\ \hline
       & -0.122 & -0.094 & -0.058 &  0.610 & -0.084 &  0.024 &  0.035   \\
117.47 & -0.094 &  0.032 & -0.039 & -0.335 &  0.024 & -0.069 &  0.006   \\
       & -0.058 & -0.039 &  0.090 &  1.107 &  0.035 &  0.006 &  0.153   \\ \hline 
\end{tabular}

\medskip

\begin{tabular}{|c|c|c|c|c|c|}   \hline
 \ \quad $Q_{12}$ \ \quad & \ \quad $Q_{23}$ \ \quad & \  \quad$Q_{13}$ \ \quad & \ \quad$Q^1_{23}$ \ \quad &
 \ \quad $Q^2_{13}$ \ \quad & \ \quad $Q^3_{12}$ \ \ \quad  \\ \hline 
  1.227  & 1.555   & 0.271    & -0.640     &  3.966     & -0.080    \\ \hline
\end{tabular}
\end{center}
\caption{
The same as Table~\ref{tbl:par-lin} for the isosceles triangle. 
\label{tbl:par-iso}
}
\end{table}

\begin{figure}[ht!]
\includegraphics[width=6cm,bb=5 50 275 220,clip]{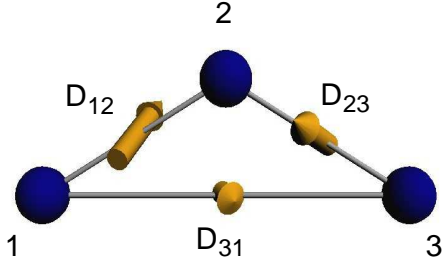}
\vskip -0.3cm
\caption{\label{fig:DM-iso} (Color online) Schematic view of the DM vectors for the isosceles Cr trimer.}
\end{figure}

The second largest contribution to the energy, namely,  the biquadratic interactions
clearly reflect the absence of global symmetry, $Q_{12} \ne Q_{23}$ and $Q_{23}^1 \ne Q^{3}_{12}$.
This asymmetry is more striking in the case of the DM vectors, see also Fig.~\ref{fig:DM-iso}. 
The good convergence of the parameters against the number of configurations seen in Fig.
 \ref{fig:DM_conv} indicates that the large asymmetry of the DM vectors 
indeed stems from the geometry of the system and not from an error of the fitting procedure.
The obtained antiferromagnetic ground state is very similar to that of the linear chain, 
see  Fig.~\ref{fig:clusters}, however, due to the absence of (mirror) symmetry the direction
of the moments is now slightly out of the line connecting sites no. 1 and 3.

\section{Summary and conclusions}
We developed a novel method in order to map the energy of supported magnetic nanoparticles obtained
from first principles calculations onto a classical spin Hamiltonian.
As a first application we determined the spin-interactions for three different Cr trimers 
deposited on a Au(111) surface.
First we calculated the electronic structure of the Cr trimers by means of
a fully relativistic Green's function embedding method.
We obtained magnetic moments of the Cr atoms in very good agreement with x-ray magnetic
circular dichroism (XMCD) measurements \cite{Ohresser} and with other first principles calculations 
\cite{bergman1,bergman2}.
The relativistic treatment of the electronic structure was inevitably
necessary to properly account for spin-orbit coupling giving rise to
tensorial exchange interactions and magnetic anisotropies influencing 
the formation of non-collinear ground states as shown in case of the equilateral trimer.

In terms of a least square fit procedure, the most general 
second-order spin-interactions as well as SU(2) invariant fourth-order terms were then
fitted serving the best approximation to the energies of a large number of random  
magnetic configurations of the trimers. 
We have shown that the inclusion of fourth-order terms
into the spin-model largely enhanced the accuracy of the mapping. 
A particular advantage of the least square fit applied in this work is that it is 
universally applicable as it does not rely on 
any symmetry restrictions on the model. 
Moreover, the spin Hamiltonian can be extended to an arbitrary order of the spin-interactions.

The magnetic ground-state of the trimers were found as the solution of the 
Landau-Lifshitz-Gilbert equations. 
In case of an equilateral Cr trimer we explored that the DM interactions lifted the degeneracy of the 
SU(2) invariant 120$^\circ$ N\'eel states with different chirality. 
On the contrary, for the linear and the isosceles Cr trimers 
we obtained collinear antiferromagnetic ground states.
An issue of choosing the reference state inherent to methods based on the magnetic force theorem
was, however, addressed in context to the equilateral Cr trimer. 
This freedom of the method 
might cause an ambiguity in determining the magnetic ground state of systems
exhibiting metastable states close to the ground state.
To overcome this problem we proposed to use the 'true' ground state obtained from
ab-initio spin dynamics calculations~\cite{fpsd} as reference, since 
the corresponding spin-model proved to be consistent with the 'parent' ground state.

The present method can be regarded as a very accurate tool in finding the magnetic 
ground state of small supported clusters providing also a clear insight into the role of 
different interactions on the formation of the magnetic ground state.
As a prospect for the future, the LLG equations will be used to study low-energy spin-excitations
of nanoparticles and the method can also be extended to include thermal 
spin-fluctuations.~\cite{antropov,privcomm}

%%%%%%%%%%%%%%%%%%%%%%%%%%%%%%%%%%%
%         ACKNOWLEDGEMENTS        %
%%%%%%%%%%%%%%%%%%%%%%%%%%%%%%%%%%%
\section{Acknowledgments}
The authors thank G. Zar\'and for fruitful discussions.
Financial support to this work was provided by the Hungarian National Scientific
Research Foundation (contract no. OTKA T068312, F68726 and NF061726).

%%%%%%%%%%%%%%%%%%%%%%%%%%%%%%%%%%%
%           BIBLIOGRAPHY          %
%%%%%%%%%%%%%%%%%%%%%%%%%%%%%%%%%%%

\end{document}